\documentclass[prb,twocolumn,showpacs,superscriptaddress,preprintnumbers,amsmath,amssymb]{revtex4}

\usepackage{graphicx}% Include figure files
\usepackage{amssymb,amsmath}
\usepackage{dcolumn}% Align table columns on decimal point
\usepackage{bm}% bold math
\usepackage{color}
\usepackage{enumerate}
\usepackage{epstopdf}

\DeclareMathOperator{\Tr}{Tr}
\DeclareMathOperator{\Imag}{Im}
\begin{document}

\title{Continuous THz emission from dipolaritons}

\author{K. Kristinsson}
\affiliation{Division of Physics and Applied Physics, Nanyang Technological University 637371, Singapore}

\author{O. Kyriienko}
\affiliation{Division of Physics and Applied Physics, Nanyang Technological University 637371, Singapore}
\affiliation{Science Institute, University of Iceland, Dunhagi-3, IS-107, Reykjavik, Iceland}

\author{T. C. H. Liew}
\affiliation{Division of Physics and Applied Physics, Nanyang Technological University 637371, Singapore}

\author{I. A. Shelykh}
\affiliation{Division of Physics and Applied Physics, Nanyang Technological University 637371, Singapore}
\affiliation{Science Institute, University of Iceland, Dunhagi-3, IS-107, Reykjavik, Iceland}

\date{\today}

\begin{abstract}
We propose a scheme of continuous tunable THz emission based on dipolaritons --- mixtures of strongly interacting cavity photons and direct excitons, where the latter are coupled to indirect excitons via tunnelling. We investigate the property of multistability under continuous wave (CW) pumping, and the stability of the solutions. We establish the conditions of parametric instability, giving rise to oscillations in density between the direct exciton and indirect modes under CW pumping. In this way we achieve continuous and tunable emission in the THz range in a compact single-crystal device. We show that the emission frequency can be tuned in a certain range by varying an applied electric field and pumping conditions. Finally, we demonstrate the dynamic switching between different phases in our system, allowing rapid control of THz radiation.
\end{abstract}

\maketitle

\section{INTRODUCTION}

The development of methods for generation of continuous radiation in the terahertz range (0.3-3 THz) is currently an important physical challenge.\cite{Siegel} Terahertz radiation has a wide area of application, ranging from astronomy \cite{MilesNATO} and nondestructive spectroscopy \cite{spectroscopy} to security and medical imaging.
Various families of devices for THz radiation generation have been theoretically proposed and experimentally tested. They include emitters based on the solid state Gunn diode \cite{Gunn} and the free-electron laser,\cite{Gold} which allow for high output power of the radiation, while suffering from a bulky size. Microscopic devices being developed for THz generation include semiconductor structures excited with femtosecond laser pulses, where oscillations of charge density induce terahertz emission by classical dipoles.\cite{femtosecond} However, the tunability of this source and its output power are highly restricted. Another microscopic emitter, proposed theoretically in 1971,\cite{SovietQCL} generates coherent THz radiation through repeated intersubband transitions across stacked quantum wells. These quantum cascade lasers (QCLs), realized experimentally in 1994,\cite{ScienceQCL} achieve high efficiency and relatively high output power (up to 600 mW),\cite{IEEEQCL} and a typical frequency of emission lying close to the upper bound of the THz range. A limiting factor of QCLs is the cryogenic temperature at which they operate. Thus, the lack of universal THz emitters with high power, efficiency, small size, which are easily-tunable and operate at relatively high temperatures stimulates the search for sources based on other operating principles.

\textit{Polaritonics} --- an area of physics which combines condensed matter physics and quantum optics \cite{KavokinBook,PolaritonDevices} --- offers new possibilities. For instance, a THz emitter based on transitions between upper and lower polariton branches was considered in Refs. [\onlinecite{KVKavokin, bistableTransition2, Savenko}]. Another possible scheme exploits the THz range transition between 2p and 1s states of the exciton, the former being a dark state and the latter coupled to the cavity mode.\cite{1S2P} A recent study suggests a bosonic QCL, with multiple THz photon emission from transitions between energy levels of exciton-polaritons confined in a parabolic potential.\cite{bosonicQCL}

\begin{figure}[t!]
\includegraphics[width=0.48\textwidth]{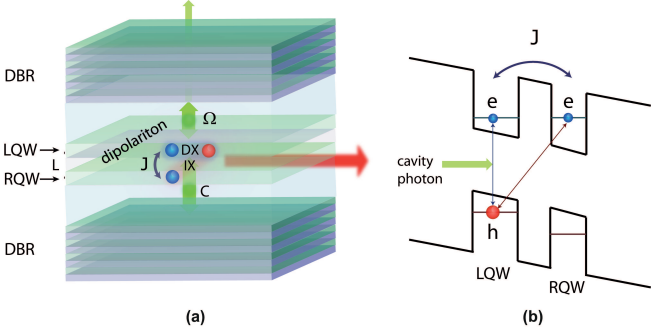}
\caption{(Color online) Sketch of the system. (a) Illustration of the optical cavity formed by two distributed Bragg reflectors (DBR) with two quantum wells (LQW and RQW) placed inside. Labels identify the cavity mode (C), direct exciton (DX) and the indirect exciton (IX), which are coupled by $\Omega$ and $J$ coupling constants and form dipolariton modes. The red arrow shows the THz radiation emitted by an oscillating dipole formed by dipolaritons. (b) Band diagram of the double quantum well tilted by applied electric field. The left QW is coupled to the cavity mode. Electron energy levels are tuned to resonance, with hopping constant $J$ between wells.}
\label{fig:sketch}
\end{figure}

This paper builds on the subject of a recent proposal to use a dipolariton system to generate pulses of THz radiation through oscillations in density between spatially direct and indirect excitons.\cite{dipolaritonTHz} The dipolariton system consists of a semiconductor microcavity with a double quantum well (QW) embedded in the center \cite{ChristmannAPL,Cristofolini} [see sketch in Fig. \ref{fig:sketch}(a)]. In a high quality cavity one can achieve strong light-matter coupling between cavity photon mode (C) and direct exciton (DX) in the left QW [Fig. \ref{fig:sketch}(b)]. In the same time, tuning electron levels of the left quantum well (LQW) and right quantum well (RQW) into resonance, one can achieve the strong tunnelling coupling between a direct exciton and a spatially indirect exciton (IX) formed by an electron in the LQW and a hole in the RQW.\cite{indirect1976,EtuningPhysRev,EtuningNature} These strong couplings between three initial modes leads to the appearance of new eigenmodes of the system, which represent three linear superpositions of the cavity photon (C), direct exciton (DX) and indirect exciton (IX) modes. They are called the upper dipolariton (UP), the middle dipolariton (MP) and the lower dipolariton (LP).

In this paper we show that accounting for nonlinear effects arising from exciton-exciton interactions can qualitatively change the behavior of the system and allow achievement of stable continuous THz emission with tunable properties. Nonlinearities are known to give rise to bistability \cite{Baas,Gippius1,Whittaker} when a single mode is excited with a coherent pump slightly above resonance, and multistability \cite{Gippius2,Paraiso} in configurations where additional states are available. Aside from the possibility of switching between different stable states,\cite{Shelykh,Schumacher,Adrados,Giorgi} parametric instabilities can achieve periodic oscillations in particle densities.\cite{Sarchi,Saito} Here, we first investigate the property of multistability under continuous wave (CW) pumping, and the stability of the solutions. Next, we establish the conditions of parametric instability, giving rise to oscillations in density between the DX and IX modes. In this way we achieve continuous and tunable emission in the THz range, in a compact single-crystal device. We show that the emission frequency can be tuned in a certain range by an applied electric field. Finally, we demonstrate the dynamic switching between different phases in our system, allowing rapid control of THz radiation.
\section{THE MODEL}
The system is represented by two coupled QWs placed inside an optical microcavity (Fig. \ref{fig:sketch}). For a coherently pumped cavity mode and resonant coupling between QWs, separated by a thin barrier, the system hosts dipolaritons --- strongly mixed modes formed by cavity photon, direct exciton and indirect exciton modes.
%The studied system is represented by two coupled QWs placed inside an optical microcavity. The left quantum well (labelled LQW in Fig. \ref{fig:sketch}) is located in the node of the resonator with cavity mode tuned to direct exciton energy. Therefore, coherent pumping of the optical cavity mode excites direct excitons in the LQW, with C and DX modes being strongly coupled. The right quantum well possesses larger energy gap and does not interact with cavity mode directly. However, due to the thin barrier between QWs and applied electric field the electron tunnelling between electrons in LQW and RQW becomes possible. For large applied voltage the system the spatially indirect exciton state is preferable, while for resonant field when DX and IX energies coincide modes are strongly coupled.
We use a generic Hamiltonian of the coupled system:
\begin{align}
\notag \hat{\mathcal H} =& \hbar \omega_C \hat a^\dagger \hat a + \hbar \omega_{DX} \hat b^\dagger \hat b + \hbar \omega_{IX} \hat c^\dagger \hat c + \frac{\hbar\Omega}{2} (\hat a^\dagger \hat b + \hat b^\dagger \hat a) \\
\notag &- \frac{\hbar J}{2}(\hat b^\dagger \hat c + \hat c^\dagger \hat b) + \frac{V_{DD}}{2} \hat b^\dagger \hat b^\dagger \hat b \hat b + \frac{V_{II}}{2} \hat c^\dagger \hat c^\dagger \hat c \hat c \\
&+ V_{DI}\text{ } \hat b^\dagger \hat c^\dagger \hat b \hat c + P(t) \hat a^\dagger + P(t)^* \hat a,
\label{hamiltonian}
\end{align}
where $\hat a^\dagger$, $\hat b^\dagger$ and $\hat c^\dagger$ are creation operators of cavity photons, direct excitons, and indirect excitons, respectively. The first three terms in Eq. (\ref{hamiltonian}) represent the energy of the bare cavity ($\hbar \omega_C$), direct exciton ($\hbar \omega_{DX}$) and indirect exciton ($\hbar \omega_{IX}$) modes. The following two terms describe the linear coupling between modes; the first denoting the Rabi splitting between the cavity mode and the direct exciton, $\hbar \Omega$, and the second being the direct to indirect exciton tunneling rate $\hbar J$. The sixth, seventh and eighth terms introduce nonlinear interactions into the system. $V_{DD}$ and $V_{II}$ are the interaction matrix elements between pairs of direct and indirect excitons, respectively. The inter-species scattering of a direct exciton with an indirect exciton is described by the matrix element $V_{DI}$. The two last terms correspond to coherent pumping of the cavity mode with intensity $|P(t)|^2$. Under CW pumping the time dependent pumping rate can be written as $P(t) = P_0 e^{-i\omega_p t}$, where $\hbar \omega_p$ is the energy of the pump and $P_{0}$ is its constant in time amplitude.

Equations of motion for the macroscopic order parameters defined as expectation values of annihilation operators $\langle \hat a_i \rangle = \text{Tr}\{\hat \rho \hat a_i\}$, $i=C,DX,IX$, can be obtained using the Heisenberg  equations of motion for operators $\hat{a},\hat{b},\hat{c}$ and applying the mean field approximation $\langle {\hat a_i} {\hat a_j} \cdots {\hat a_k} \rangle \approx \langle {\hat a_i} \rangle \langle {\hat a_j} \rangle \cdots \langle {\hat a_k} \rangle$. Additionally we perform the change of variables $\hat a_i \rightarrow e^{-i\omega_C t}\hat a_i$. The resulting system of equations reads:
\begin{align}
\frac{\partial \langle {\hat a} \rangle}{\partial t} =& -i \frac{\Omega}{2} \langle \hat b\rangle -\frac{1}{2\tau_C} \langle \hat a \rangle - i \tilde{P}(t) ,
\label{Ceqn}\\
\notag
\frac{\partial \langle \hat b \rangle} {\partial t} =& i\delta_\Omega \langle \hat b \rangle -i\frac \Omega 2 \langle \hat a \rangle + i \frac J 2 \langle \hat c \rangle - \frac 1 {2\tau_{DX}} \langle \hat b \rangle \\
&- \frac i \hbar (V_{DD} |\langle \hat b \rangle|^2 + V_{DI} |\langle \hat c \rangle|^2) \langle \hat b \rangle ,
\label{DXeqn}\\
\notag
\frac{\partial \langle \hat c \rangle} {\partial t} =& i(\delta_\Omega - \delta_J)\langle \hat c \rangle + i \frac J 2 \langle \hat b \rangle - \frac 1 {2\tau_{IX}} \langle \hat c \rangle \\
&- \frac i \hbar (V_{II}|\langle \hat c \rangle|^2 + V_{DI}|\langle \hat b \rangle|^2)\langle \hat c \rangle,
\label{IXeqn}
\end{align}
where we introduced the lifetimes of the modes $\tau_C = 5$ ps, $\tau_{DX} = 1$ ns and $\tau_{IX}=100$ ns. Explicit references to mode energies have been removed by defining relative energies $\delta_{\Omega} = \omega_C - \omega_{DX}$ and $\delta_J = \omega_{IX} - \omega_{DX}$. The pumping term now reads $\tilde P (t) = e^{i\omega_Ct}P(t)/\hbar$. Under CW pumping we have $\tilde P (t) = \tilde P _0 e^{-i \Delta_pt}$, where $\tilde P _0 = P_0/\hbar$, and $\Delta_p = \omega_p - \omega_C$ is the relative pumping frequency.

The conditions investigated in this paper correspond to high occupation numbers of the modes for which the first order mean field approximation is applicable. To test this we also derived dynamic equations written for higher order mean field theory [see Appendix A]. Numerical solution showed no change in the results, confirming the accuracy of Eqns. (\ref{Ceqn})--(\ref{IXeqn}).

In addition to dynamics we can study the steady-state properties of the system assuming a CW pump with tunable energy and intensity. It was shown for the case of a pumped mode with nonlinearity present in the system that several solutions for the occupation numbers of the modes as a function of pumping intensity are possible.\cite{oneModeNonl,noLinCoupl,noCrossNonl,differentPump} Moreover, in the case of two coupled modes the system can exhibit either stable solutions or parametrically unstable solutions characterized by continuous oscillations of the mode occupation numbers.\cite{noCrossNonl}

The stationary solutions can be found using the ansatz for the modes \cite{noCrossNonl}
\begin{equation}
\langle \hat a_i\rangle=e^{-i\Delta_p t}\psi_i^0 ,~~~i=C,DX,IX ,
\end{equation}
which corresponds to harmonic oscillations with the frequency of the CW pump. Inserting this into Eqns. (\ref{Ceqn})--(\ref{IXeqn}) and eliminating the non-interacting photonic mode $\psi_C$, we obtain a coupled pair of equations for the occupation numbers of the exciton modes:
\begin{align}
&(-E_1 - i\gamma_1 + V_{DD}|\psi_{DX}^0|^2 + V_{DI} |\psi_{IX}^0|^2 )\psi_{DX}^0 \notag \\
&- \frac 1 2 \hbar J \psi_{IX}^0 + P_1 = 0, \label{stationary1}\\
&(-E_2 - i\gamma_2 + V_{DI} |\psi_{DX}^0|^2 + V_{II} |\psi_{IX}^0|^2 )\psi_{IX}^0 \notag \\
&- \frac 1 2 \hbar J \psi_{DX}^0 =0, \label{stationary2}
\end{align}
where $E_i,\gamma_i,P_1$, $i=1,2$ are dressed energies, decay rates and pumping strength, expressed as
\begin{align*}
E_1 &= \hbar \Delta_p + \hbar \delta_\Omega - \frac {(\hbar \Omega)^2 \hbar \Delta_p} {(2\hbar \Delta_p)^2 + \gamma_C^2}, \\
E_2 &= \hbar \Delta_p + \hbar \delta_\Omega - \hbar \delta_J , \\
\gamma_1 &= \frac {\gamma_{DX}} 2 + \frac {(\hbar \Omega)^2} {8(\hbar \Delta_p)^2 + 2 \gamma_C^2} \gamma_C ,\\
\gamma_2 &= \frac {\gamma_{IX}} 2 ,\\
|P_1| &= \frac {\hbar \Omega}{\sqrt{(2\hbar \Delta_p)^2 + \gamma_C^2}}.
\end{align*}
Equations (\ref{stationary1}) and (\ref{stationary2}) represent a system of nonlinear equations for the IX and DX occupation numbers with multiple solutions in certain ranges of the pumping intensity.

The stability analysis of solutions can be performed similarly to the method used in Ref. [\onlinecite{noCrossNonl}]. Eqns. (\ref{stationary1}), (\ref{stationary2}) can be recast as dynamic equations
\begin{align}
i\hbar \partial_t \psi_{DX}= &(\hbar\Delta_p -E_1 - i\gamma_1 + V_{DD}|\psi_{DX}|^2 + \notag \\ 
&+ V_{DI} |\psi_{IX}|^2 )\psi_{DX} - \frac 1 2 \hbar J \psi_{IX} + P_1(t), \label{dynamic1} \\
i\hbar \partial_t \psi_{IX}= &(\hbar\Delta_p -E_2 - i\gamma_2 + V_{DI} |\psi_{DX}|^2 + \notag \\
&+ V_{II} |\psi_{IX}|^2 )\psi_{IX} - \frac 1 2 \hbar J \psi_{DX}, \label{dynamic2}
\end{align}
with CW solutions written in the form $\psi_i(t) = e^{-i\Delta_p t}\psi_i^0$. To determine their stability let us explore the dynamics of a trial wave function 
\begin{equation}
\psi_i(t) = e^{-i\Delta_p t}\big(\psi_i^0 + \delta\psi_i(t)\big),~~~i=DX,IX, 
\label{deviation}
\end{equation}
where $\delta\psi_i(t)$ is a small deviation from the stationary solution. We insert (\ref{deviation}) into Eqns. (\ref{dynamic1})--(\ref{dynamic2}), use Eqns. (\ref{stationary1})--(\ref{stationary2}) to eliminate the pumping term $P_1(t)$, and neglect terms of second order and higher in $\delta \psi_i$. Finally, we get dynamic equations for deviations $\delta\psi_i(t)$
\begin{widetext}
\begin{align}
i\hbar \partial_t \delta\psi_{D} &= \Big[-E_1 -i\gamma_1 +2V_{DD}|\psi_{D}^0|^2 +V_{DI}|\psi_{I}^0|^2 \Big]\delta\psi_{D} + V_{DD}{\psi_{D}^0}^2 \delta\psi_{D}^* + V_{DI} \psi_{D}^0\psi_{I}^0\delta\psi_{I}^* + {\Big(V_{DI}\psi_{I}^0}^* \psi_{D}^0 -\frac 1 2 \hbar J \Big)\delta\psi_{I}, \\
i\hbar \partial_t \delta\psi_{I} &= \Big[-E_2 -i\gamma_2 +2V_{II}|\psi_{I}^0|^2 +V_{DI}|\psi_{D}^0|^2 \Big]\delta\psi_{I} + V_{II}{\psi_{I}^0}^2 \delta\psi_{I}^* + V_{DI} \psi_{I}^0\psi_{D}^0\delta\psi_{D} ^* + \Big(V_{DI}{\psi_{D}^0}^* \psi_{I}^0 -\frac 1 2 \hbar J \Big)\delta\psi_{D},
\end{align}
\end{widetext}
where index D (I) denotes direct (indirect) exciton.

We now assume that the time dependence of the deviation can be expressed as $\delta\psi_{i} = u_i e^{-iEt/\hbar} + v_i e^{iE^* t/\hbar}$. Inserting this ansatz in above equations yields an eigenvalue problem $\mathcal{M} \delta\Psi = E \delta\Psi$ for $\delta\Psi := (u_D,v_D,u_I,v_I)$, where
\begin{widetext}
\begin{equation}
\mathcal{M} =
 \begin{pmatrix}
  -E_1 -i\gamma_1 & V_{DD}{\psi_D^0} ^2 & V_{DI}\psi_D^0 {\psi_I^0}^* - \frac 1 2 \hbar J & V_{DI} \psi_D^0 \psi_I^0 \\
+ 2V_{DD}|\psi_{D}^0|^2 +V_{DI}|\psi_{I}^0|^2  & & & \\
 & & & \\
  -V_{DD}({\psi_D^0}^*)^2 & E_1 -i\gamma_1 & -V_{DI}{\psi_D^0}^* {\psi_I^0}^* & -V_{DI} {\psi_D^0}^* \psi_I^0 + \frac 1 2 \hbar J \\
 & -2V_{DD}|\psi_{D}^0|^2 -V_{DI}|\psi_{I}^0|^2 & & \\
 & & & \\
  V_{DI} {\psi_D^0}^* \psi_I^0 -\frac 1 2 \hbar J & V_{DI}\psi_D^0 \psi_I^0 & -E_2 -i\gamma_2 & V_{II}{\psi_I^0}^2  \\
 & & +2V_{II}|\psi_{I}^0|^2 +V_{DI}|\psi_{D}^0|^2 & \\
 & & & \\
  -V_{DI} {\psi_D^0}^* {\psi_I^0}^* & -V_{DI}\psi_D^0 {\psi_I^0}^* + \frac 1 2 \hbar J & -V_{II} ({\psi_I^0}^*)^2  & E_2 -i\gamma_2 \\
 & & & -2V_{II}|\psi_{I}^0|^2 -V_{DI}|\psi_{D}^0|^2
 \end{pmatrix}
\end{equation}
\end{widetext}
For each stationary state $\psi_D^0$, $\psi_I^0$ there are four eigenvalues $E_\alpha$. The stability of the solutions can be determined from the sign of the eigenvalues. For instance, keeping in mind the form of the deviation, we see it decays in time if all imaginary parts are negative, $\Imag(E_\alpha)<0 ~ \forall \alpha$. The solution then is stable with regards to small perturbations. If the imaginary part of one eigenvalue is non-negative, $\Imag(E_\alpha)\geq 0$, the solution is unstable. If additionally Re$(E_\alpha)>0$, the solution is parametrically unstable.\cite{noCrossNonl}
\begin{figure}[t!]
\includegraphics[width=0.48\textwidth]{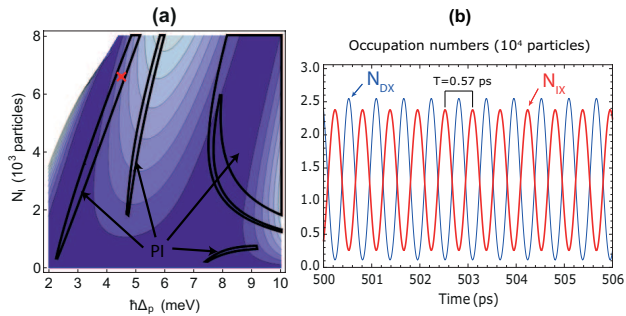}
\caption{(Color online) (a) Phase diagram plotted for indirect exciton population numbers on the vertical axis and the pump energy on the horizontal axis. Enclosed regions indicate parametric instability (PI). Elsewhere the system is stable. Lighter colors indicate higher pumping strengths are needed to reach relevant population numbers. We can identify the resonance with the MP mode and the UP mode as dark regions. Both modes shift higher in energy with higher population numbers. The white region in the upper left corner indicates population numbers that can not be reached at this pumping energy. Pumping conditions used in (b) are identified with a red cross. (b) Time evolution of direct and indirect exciton populations in the region of parametric instability. At large times populations oscillate harmonically.}
\label{fig:phase_diagram}
\end{figure}
\begin{figure*}
\includegraphics[width=0.92\textwidth]{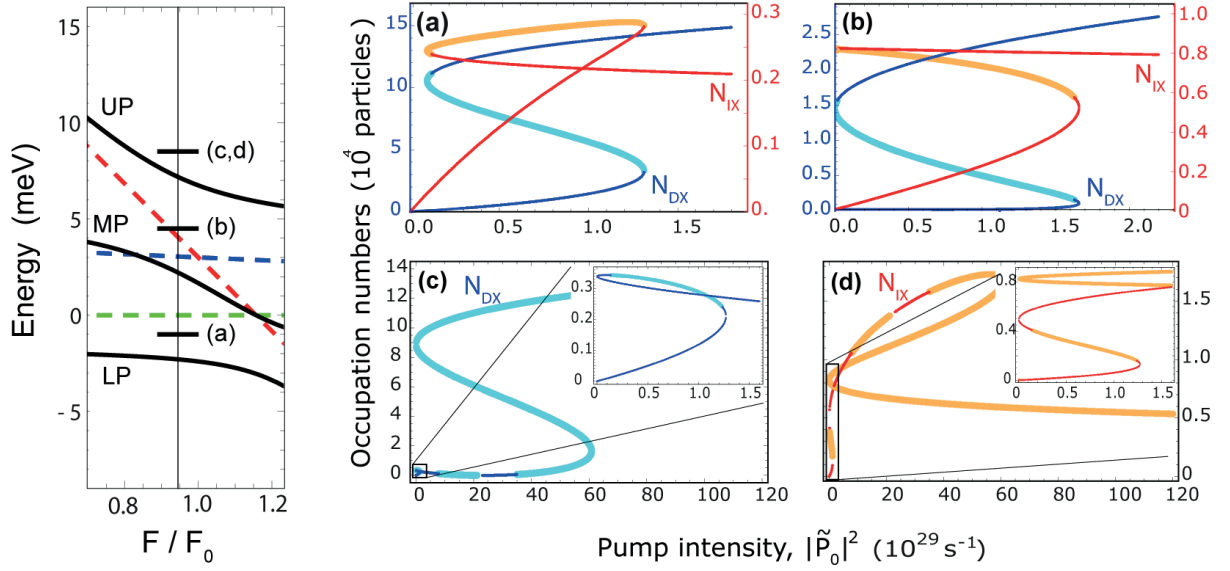}
\caption{(Color online) The leftmost plot shows the energy of the modes as a function of the applied electric field and indicates pumping energies. The dashed lines indicate the cavity (green), direct exciton (blue) and indirect exciton (red) modes, and full lines indicate lower (LP), middle (MP) and upper polariton (UP). Pumping energies are indicated with short horizontal lines, and letters refer to relevant bistability curves. The four tiled plots show multistability under CW pumping. Blue curves correspond to direct exciton occupation numbers, and red curves to indirect exciton occupation numbers. Thin curves indicate stability of the population numbers, while thick curves indicate the presence of parametric instability. (a) The pumping energy $\hbar \Delta_p = -1$ meV is slightly higher than LP energy. Here, the bistability is induced by the blueshift of the LP mode. The DX number (left axis) is much larger than the IX number (right axis), reflecting the small IX fraction in the LP mode. (b) The pumping energy $\hbar \Delta_p = 4.5$ meV is slightly higher than the MP energy, and bistability is induced by the blueshift of the MP mode. The IX number is of the same order of magnitude as the DX number, reflecting their almost equal fractions in the MP mode. This regime facilitates large oscillations in dipole moment under parametric instability. (c, d) The pumping energy $\hbar \Delta = 8.5$ meV is slightly higher than the UP energy. The insets show mulstistability governed by the blueshift of the UP mode, analogous to plots (a) and (b). The main plots shows multistability originating in the blueshift of both the UP mode at low pumping strengths, and from the MP mode. For the latter the pump intensities needed are much higher than in previous cases, as the pump-MP mode detuning is much larger.}
\label{fig:bistability_curves}
\end{figure*}

\section{RESULTS}

We model the system based on GaAs/AlGaAs quantum well parameters, with linear coupling parameters chosen as $\hbar J = \hbar \Omega = 6$ meV.\cite{Cristofolini} The DX-DX interaction constant can be estimated as $V_{DD} = 6E_b a_B^2 /S$,\cite{Tassone} where the direct exciton Bohr radius and binding energy are $a_B = 10$ nm and $E_b = 8$ meV, respectively. Here $S=100$ $\mu$m$^2$ is the laser excitation area. The IX-IX scattering constant was taken from Ref. [\onlinecite{IXIXint}], with the QW separation taken as $L=12$ nm (4 nm tunnelling barrier). The derivation of the DX-IX interaction constant is shown in Appendix B. Relative energies of the modes were chosen as $\hbar \delta_{\Omega} = -3$ meV and $\hbar \delta_J = 1$ meV, the latter corresponding to an applied electric field of magnitude $F = 0.945 F_0$, where $F_0=12.5$ kV/cm is the field strength at which the DX and IX modes are resonant.\cite{DXIXresonance}

Performing numerically the stability analysis of solutions of Eqns. (\ref{stationary1}) and (\ref{stationary2}) allows us to plot a phase diagram of the system shown in Fig. \ref{fig:phase_diagram}(a). For certain values of pumping energy $\hbar \Delta_{p}$ and pump intensity, connected to the number of indirect excitons $N_{IX}$, solutions that are not stable are possible. In these regions we can have parametric instability (PI).

Multistability curves plotted for pumping energies $\hbar \Delta_p=-1,4.5,8.5$ meV are shown in Fig. \ref{fig:bistability_curves}(a), (b) and (c,d), respectively. Stable steady state solutions are shown with narrow darker lines, and thicker lines indicate parametric instability. One can see that the DX populations are reminiscent of the conventional one mode polariton system bistability,\cite{Baas,Gippius1,Whittaker} with regions of instability now replaced with parametric instability.
This implies that population numbers do not decay to stable steady state solutions, but oscillate continuously in time. Solving numerically Eqns. (\ref{Ceqn})--(\ref{IXeqn}) we show long-standing beats of IX and DX occupation numbers under CW pump (Fig. \ref{fig:phase_diagram}(b)). Note that this behavior of the system is only possible due to the presence of interactions, while without accounting for nonlinearities beats of IX-DX density quickly decay in time.\cite{dipolaritonTHz}

\begin{figure}[b!]
\includegraphics[width=0.45\textwidth]{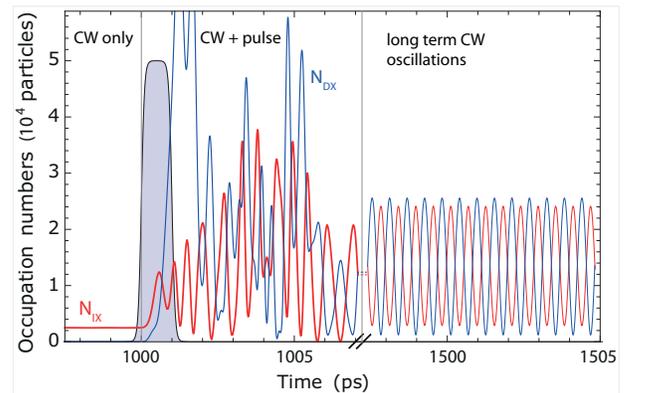}
\caption{(Color online) Dynamical switching of the system. The system is stable on the lower branch, when at $t=1000$ ps an additional optical pusle is applied. This drives the system to parametric instability, resulting at large times in harmonic oscillations of IX and DX occupation numbers. The system has been switched to the THz radiating state.}
\label{fig:pulseSwitch}
\end{figure}

The pumping energy was now fixed to $\hbar \Delta_p=4.5$ meV [case (b) in Fig. \ref{fig:bistability_curves}] and the CW pump intensity was linearly turned on to $|{\tilde P}_0|^2 = 1.1\cdot 10^{29}$ s$^{-1}$. For long turning on times ($>5$ ps) the system remains stable on the lower branches of bistability, with $N_{IX} = 2500$, and $N_{DX} \ll N_{IX}$. For short turning on times ($<2$ ps) the system is excited to the parametrically unstable middle branch, resulting at large times in harmonic oscillations of the population numbers. The system can also be switched from the stable behaviour to the parametrically unstable one by applying additionally to CW pumping a short laser pulse, as shown in Fig. \ref{fig:pulseSwitch}. This represents the high degree of control of the proposed system.

\begin{figure}[t!]
\includegraphics[width=0.4\textwidth]{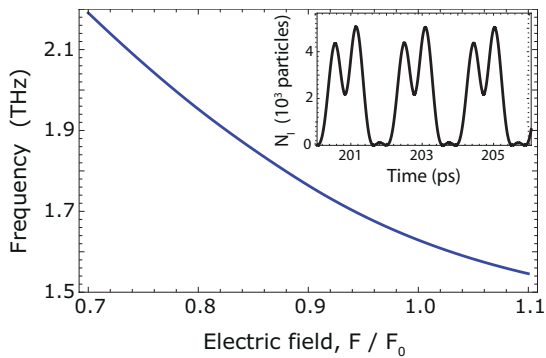}
\caption{(Color online) Frequency of oscillations of indirect exciton occupation numbers plotted as a function of the applied electric field, for cavity mode detuning $\delta_{\Omega} =-3$ meV and pumping energy $\Delta_p = 4.5$ meV. Inset shows anharmonic oscillations in indirect exciton numbers for electric field $F=1.2F_0$.}
\label{fig:freqF}
\end{figure}

Furthermore, we study the behavior of the system subjected to the variation of the detuning $\delta_J$ between IX and DX modes, controlled by the applied field $F$. The energy and intensity of the pump are kept constant. This change alters the frequency of particle numbers oscillations. The dependence of the frequency on the applied field is shown in Fig. \ref{fig:freqF}, with frequency spanning from 1.5 to 2.2 THz. For the values of $F$ past $1.1F_0$ oscillations become strongly anharmonic, shown in the inset in Fig. \ref{fig:freqF}, which means that the system demonstrates pronounced multi-mode emission. When the electric field is decreased past the limits of Fig. \ref{fig:freqF}, the time to reach harmonic oscillations increases rapidly, limiting the usefulness of this range.
\section{DISCUSSION}
We have shown that exploiting parametric instability we can achieve harmonic oscillations in the indirect exciton population under CW pumping conditions. An important property of spatially indirect excitons is a non-zero dipole moment directed in the growth direction of the microcavity. The single indirect exciton carries dipole moment equal to $d_0 = eL$, where $L$ is the QW separation, while the total dipole moment of the system can be calculated as $D(t) = d_0 N_{IX}(t)$. In the case of the oscillating population numbers, this can be written:
\begin{equation}
D(t) = d_0 N_{IX}^0 \cos(\omega t /2)^2 ,
\end{equation}
where $\omega= 2 \pi \nu$ is the angular frequency of the IX population oscillations, and $N_{IX}^0$ is an amplitude of the IX occupation number modulation. Thus, oscillations of the IX density induce oscillations of the total dipole moment of the system with frequency lying in the terahertz range. This results in the emission of THz radiation by the array of classical dipoles formed from dipolaritons excited over an area in the microcavity plane.
To estimate the power of emitted THz radiation, one can treat the system as a classical dipole antenna, which has a far field radiation intensity given by \cite{antenna}
\begin{equation}
I_0 = \frac {{\ddot D}_{RMS}^2} {6\pi \epsilon_0 c^3}=\frac {(N_{IX}^0 d_0 \omega^2)^2} {3\pi \epsilon_0 c^3},
\end{equation}
where $\epsilon_0$ is the vacuum permittivity, and $c$ is the speed of light. The radiation intensity is directionally dependent, as $I(\theta) \propto \sin^2(\theta)$, where $\theta$ is the angle of emission with respect to the microcavity growth axis. Choosing the applied field as $F=0.945 F_0$, the frequency of oscillations is 1.75 THz, and the maximum IX number is $N_{IX}=2.4\cdot 10^4$. The total emitted power is then $I_0 \approx 14$ nW.

It is important to note that the intensity is quadratic in the indirect exciton number. This can be explained by the phenomenon of superradiance \cite{Dicke,superradianceNature} --- coherence of the oscillations in the quantum system causes the emitted power to increase superlinearly in the number of oscillators. The typical area over which this coherence can be realized is given by the pumping spot diameter. For example, a diameter of 60 $\mu$m would give an emitted power $I \approx 11$ $\mu$W. The emission intensity can be further enhanced by growing additional stacks of double QWs, and by placing the system inside a supplemental THz cavity, allowing stimulated emission.

In a polariton system with GaAs/AlGaAs quantum wells, operating temperature are limited to temperatures of strong coupling observation, typically around 70 K.\cite{Grosso} However, one can expect that with nitride III-V compounds the operating temperature of the dipolariton emitter can be increased. All this serves to give this scheme competitive characteristics compared to other solid-state THz sources.

\section{CONCLUSION}

We have shown that nonlinear interactions in a dipolariton system give rise to multistability effects. In particular, for certain values of pumping parameters the parametric instability between IX and DX modes occurs. This results in continuous oscillations of the spatially indirect exciton occupation number under CW pumping. The frequency of these oscillations is in the THz range and can be tuned by an applied electric field. Depending on the parameters, the resulting THz radiation by the array of classical dipoles represents a continuous single-mode or multi-mode THz laser, with power output and efficiency expected to be improved over existing solid state THz emitters. We have also shown rapid switching of the THz emission controlled by an additional short laser pulse.\\

\begin{acknowledgements}
We would like to thank A. V. Kavokin for useful discussions. This work has been supported by FP7 IRSES project ``POLATER'' and AcRF Tier 1 project ``Polaritonics for Novel Device Applications''. O. K. acknowledges the support from Eimskip Fund.
\end{acknowledgements}

\appendix

\section{Equations of motion in higher order mean field theory}
In the following section we derive dynamical equations for the relevant expectation values to higher order mean field theory using the density matrix formalism. The higher order equations are then shown not to yield observable increase in accuracy.

The total Hamiltonian of the system can be written as a sum of terms describing coherent and incoherent processes, $\hat H = \hat H_{coh} + \hat H_{dec}$. The master equation for the density matrix $\rho$ reads:
\begin{equation}
i\hbar \frac{\partial \rho}{\partial t}=[\hat H_{coh},\rho]+\hat{\mathcal{L}}\rho, 
\end{equation}
where $\hat{\mathcal L}$ is the Lindblad superoperator.\cite{Savenko,1S2P} It accounts for dissipation processes such as non-radiative recombination of excitons, and leakage of cavity mode photons due to the non-zero transmission resonator mirrors. The explicit form of the term is
\begin{align}
\notag \hat{\mathcal L} = &\frac {i\gamma_C} 2 (2 a\rho a^{\dagger} -a^{\dagger} a\rho - \rho a^{\dagger} a) +\frac {i\gamma_{DX}} 2 (2 b\rho b^{\dagger} - b^{\dagger} b\rho \\
& - \rho b^{\dagger} b) +\frac {i\gamma_{IX}} 2 (2 c\rho c^{\dagger} -c^{\dagger} c\rho - \rho c^{\dagger} c).
\label{Lindblad}
\end{align}
Here $a$ is the annihilation operator for the cavity mode (C), $b$ is the annihilation operator of the direct exciton (DX), and $c$ that of the indirect exciton (IX), and we omit hats of operators for simplicity. The coefficients $\gamma_i = \hbar /{\tau_i}$ are the decay rates of those modes ($i=C,DX,IX$).

The equation of motion for the expectation value of an operator $A$ reads
\begin{equation}
i\hbar \frac{\partial \langle A \rangle}{\partial t} = i\hbar \frac \partial {\partial t} \Tr\{ A \rho\}= \langle [A,\hat H_{coh}] \rangle + \Tr\{A\hat{\mathcal L}\rho\}.
\label{OperatorEquation}
\end{equation}
Equations of motion can be derived using simple operator algebra, and applying the mean field approximation. For illustration purposes, let us derive the terms corresponding to the nonlinear interaction part of the Hamiltonian,
\begin{equation} 
H_{nonl} = \frac {V_{DD}} 2 b^\dagger b^\dagger b b + \frac {V_{II}} 2 c^\dagger c^\dagger c c + V_{DI}b^\dagger c^\dagger b c,
\end{equation}
where $V_{DD}$ is the scattering matrix element for the interaction between two direct excitons, $V_{II}$ is the indirect-indirect exciton interaction matrix element, and $V_{DI}$ is the matrix element for scattering of a direct and an indirect exciton, calculated in Appendix B.
Using the equation (\ref{OperatorEquation}) the nonlinear part of dynamics for operator $A$ reads
\begin{equation} 
\{ \partial_t \langle A\rangle \}_{nonl} = \frac 1 {i\hbar}\langle [A,H_{nonl}]\rangle  . 
\label{ExpVal}
\end{equation}
Now, all number operators of the system commute with $H_{nonl}$. Applying this equation to the particle numbers operators $N_i$, $i=C,DX,IX$ we get
\begin{equation} 
\{ \partial_t \langle N_C\rangle \}_{nonl} = \{ \partial_t \langle N_{DX}\rangle \}_{nonl} = \{ \partial_t \langle N_{IX}\rangle \}_{nonl} = 0. 
\end{equation}
Let us now apply Eq. (\ref{ExpVal}) to the correlator $\alpha = \langle a^\dagger b\rangle$:
\begin{align} \notag  \{ \partial_t \alpha \}_{nonl} =& \frac 1 {2i\hbar}\langle [a^\dagger b;V_{DD} b^\dagger b^\dagger b b + V_{II} c^\dagger c^\dagger c c + \notag \\
& +2V_{DI}b^\dagger c^\dagger b c]\rangle \notag \\
=& \frac 1 {2i\hbar}V_{DD} \langle a^\dagger [b,b^\dagger b^\dagger] b b\rangle + \frac 1 {i\hbar} V_{DI} \langle a^\dagger [b,b^\dagger] c^\dagger b c\rangle \notag \\
=&- \frac {i} {\hbar} \Big(V_{DD} \langle a^\dagger b^\dagger b b \rangle + V_{DI} \langle a^\dagger c^\dagger b c \rangle \Big) \notag \\
\approx & - \frac {i} {\hbar} \Big(V_{DD} \langle a^\dagger b\rangle \langle b^\dagger b \rangle + V_{DI} \langle a^\dagger b \rangle \langle c^\dagger  c \rangle \Big) \notag \\
=&- \frac {i} {\hbar} \Big(V_{DD} N_{DX}+ V_{DI} N_{IX}\Big)\alpha.
\end{align}
In the second to last step we used second order mean field theory to close the system of equations. Equations of motion for correlators $\beta = \langle b^\dagger c\rangle $ and $\gamma= \langle c^\dagger a\rangle $ can similarly be calculated as
\begin{align} 
\{ \partial_t \beta  \}_{nonl} &= \frac {i} {\hbar} \Big( (V_{DD}-V_{DI})N_{DX}+ (V_{DI}-V_{II})N_{IX}\Big)\beta \\
\{ \partial_t \gamma \}_{nonl} &= \frac {i} {\hbar} \Big(V_{DI} N_{DX}+ V_{II} N_{IX}\Big)\gamma.
\end{align}

Next, we derive the dynamic equations for expectation values of the annihilation operators. Inserting the operator $b$ in Eq. (\ref{ExpVal}) we get
\begin{align} \notag  \{ \partial_t \langle b \rangle \}_{nonl} =& \frac 1 {2i\hbar}\langle [b;V_{DD} b^\dagger b^\dagger b b + V_{II} c^\dagger c^\dagger c c + \\
&+ 2V_{DI}b^\dagger c^\dagger b c]\rangle \notag \\
=& \frac 1 {2i\hbar}V_{DD} \langle [b,b^\dagger b^\dagger] b b\rangle + \frac 1 {i\hbar} V_{DI} \langle [b,b^\dagger] c^\dagger b c\rangle \notag \\
=&- \frac {i} {\hbar} \Big(V_{DD} \langle b^\dagger b b \rangle + V_{DI} \langle c^\dagger b c \rangle \Big) \notag \\
\approx & - \frac {i} {\hbar} \Big(V_{DD} \langle b\rangle \langle b^\dagger b \rangle + V_{DI} \langle b \rangle \langle c^\dagger  c \rangle \Big) \notag \\
=&- \frac {i} {\hbar} \Big(V_{DD} N_{DX}+ V_{DI} N_{IX}\Big)\langle b \rangle,
\end{align}
and in the same way derive equation for $c$,
\begin{equation}
\{ \partial_t \langle c \rangle \}_{nonl} = - \frac {i} {\hbar} \Big(V_{DI} N_{DX}+ V_{II} N_{IX}\Big)\langle c \rangle. \end{equation}
Note that creation or annihilation operators for the cavity mode does not appear in $H_{nonl}$, and
\begin{equation}
\{ \partial_t \langle a \rangle \}_{nonl} = 0.
\end{equation}

\begin{figure}[b!]
\includegraphics[width=0.48\textwidth]{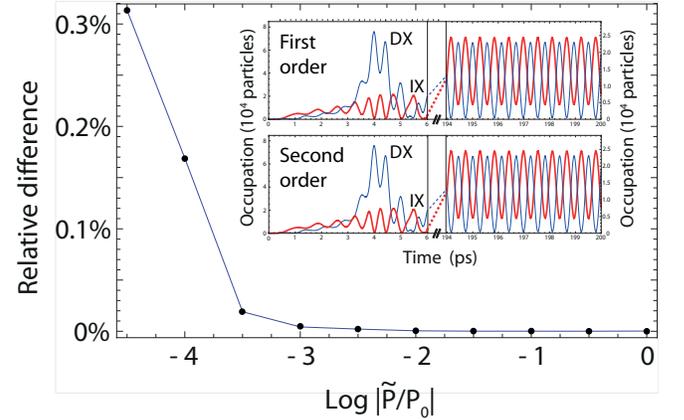}
\caption{(Color online) Relative difference between stable population numbers, obtained from equations (\ref{Ceqn})-(\ref{IXeqn}) and equations (\ref{second_order_0})-(\ref{second_order_end}), as a function of pumping strength. The scale of the pumping strength is relative to a typical strength $|P_0|^2 = 10^{29}$ s$^{-1}$. The difference is negligable at the typical pumping strength. The inset shows a comparison of solutions using the first and second order mean field theory, for a short turning on time of the CW pump. Even at large times there is no observable difference in the solutions, confirming there is no increase in accuracy. }
\label{fig:MFT_comparison}
\end{figure}

Repeating this procedure for the other terms in the Hamiltonian leads to a closed set of equations. We perform a change of variables $a_i \rightarrow e^{-i\omega_c t} a_i$, $i = C, DX, IX$. This preserves particle numbers and correlators, and removes reference to absolute mode energies. Equations can then be expressed using relative energies $\delta_\Omega = \omega_C -\omega_{DX}$, $\delta_J = \omega_{IX}-\omega_{DX}$, and read
\begin{widetext}
\begin{align}
\label{second_order_0}
&\partial_t N_C = \Omega \Imag(\alpha) - \frac 2 \hbar \Imag (\tilde{P}(t)^*\langle a \rangle ) - \frac 1 {\tau_C} N_C ,\\
&\partial_t N_{DX} = -\Omega \Imag(\alpha) - J \Imag(\beta) - \frac 1 {\tau_{DX}} N_{DX},\\
&\partial_t N_{IX} = J \Imag(\beta) - \frac 1 {\tau_{IX}} N_{IX},\\
&\partial_t \alpha = i \delta_\Omega \alpha + i \frac \Omega 2 (N_{DX} - N_C) + i \frac J 2 \gamma^* -\frac i \hbar (V_{DD} N_{DX} + V_{DI} N_{IX} ) \alpha + i \tilde{P}(t)^* \langle b \rangle - \frac 1 2 \left(\frac 1 {\tau_C} + \frac 1 {\tau_{DX}}\right) \alpha,\\
&\partial_t \beta = -i\delta_J \beta + i \frac \Omega 2 \gamma^* + i \frac J 2 (N_{DX} - N_{IX}) + \frac i \hbar \Big[ (V_{DD} - V_{DI}) N_{DX} +(V_{DI}-V_{II})N_{IX} \Big] \beta - \frac 1 2 \left( \frac 1 {\tau_{DX}} + \frac 1 {\tau_{IX}}\right) \beta ,\\
&\partial_t \gamma =i(\delta_J -\delta_\Omega) \gamma -i\frac \Omega 2 \beta^* - i \frac J 2 \alpha^* + \frac i \hbar(V_{II}N_{IX} + V_{DI}N_{DX}) \gamma -i \tilde{P}(t) \langle c \rangle^* - \frac 1 2 \left(\frac 1 {\tau_{IX}} + \frac 1 {\tau_C}  \right)\gamma ,\\
%\end{align}
%\begin{align}
&\partial_t \langle a \rangle = -i \frac \Omega 2 \langle b\rangle - i \tilde{P}(t) -\frac 1 {2\tau_C} \langle a \rangle ,\\
&\partial_t \langle b \rangle = i\delta_\Omega \langle b \rangle -i\frac \Omega 2 
\langle a \rangle + i \frac J 2 \langle c \rangle - \frac i \hbar (V_{DD} N_{DX} + V_{DI} N_{IX}) \langle b \rangle - \frac 1 {2\tau_{DX}} \langle b \rangle ,\\
&\partial_t \langle c \rangle = i(\delta_\Omega - \delta_J)\langle c \rangle + i \frac J 2 \langle b \rangle - \frac i \hbar (V_{II}N_{IX} + V_{DI}N_{DX})\langle c \rangle - \frac 1 {2\tau_{IX}} \langle c \rangle,
\label{second_order_end}
\end{align}
\end{widetext}
where $\tilde{P}(t) = e^{i\omega_C t}P(t)$.

For accuary comparison, numerical modelling was done in two ways. First the system of the nine coupled equation written for second order mean-field approximation were solved for certain conditions. Second the three first order equations for expectation values $\langle a \rangle$, $\langle b \rangle$ and $\langle c \rangle$ were solved for the same conditions, setting $N_i = |\langle a_i \rangle|^2$. The latter correspond to  Eqns. (\ref{Ceqn})--(\ref{IXeqn}) in the main text. The results of this comparison is shown in Fig. \ref{fig:MFT_comparison}. The second order theory is shown only to yield observable increase in accuracy for pumping strenghts many orders of magnitude smaller than the ones treated in the paper. For this reason the first order theory was chosen to be presented in the paper, and used for obtaining the results shown.

\section{DX-IX interaction constant}

The calculation of the matrix element of the interaction between the direct and indirect exciton follows the derivation of IX-IX interaction constant performed in Ref. [\onlinecite{IXIXint}]. However, the difference comes from difference of wave function of interacting particles corresponding to direct and indirect exciton, and distinct geometry of the Coulomb interaction between carriers. The matrix element can be written as the sum $V_{DX-IX} = V_{DX-IX}^{dir} + V_{DX-IX}^{X-exch} + V_{DX-IX}^{e-exch} + V_{DX-IX}^{h-exch}$, where
\begin{widetext}
\begin{align}
\label{Vdir}
V_{DX-IX}^{dir}&(\mathbf{Q},\mathbf{Q'},\mathbf{q})= \notag \\
&= \int d^{2}\mathbf{r_{e}} d^{2}\mathbf{r_{h}} d^{2}\mathbf{r_{e'}} d^{2}\mathbf{r_{h'}} \Psi^{DX}_{\mathbf{Q}}(\mathbf{r_{e}},\mathbf{r_{h}})^{\ast}\Psi^{IX}_{\mathbf{Q'}}(\mathbf{r_{e'}},\mathbf{r_{h'}})^{\ast}V_{I}(\mathbf{r_{e}},\mathbf{r_{h}},\mathbf{r_{e'}},\mathbf{r_{h'}})\Psi^{DX}_{\mathbf{Q+q}}(\mathbf{r_{e}},\mathbf{r_{h}})\Psi^{IX}_{\mathbf{Q'-q}}(\mathbf{r_{e'}},\mathbf{r_{h'}}),\\
\label{Vx}
V_{DX-IX}^{X-exch}&(\mathbf{Q},\mathbf{Q'},\mathbf{q})=\notag \\
&=   \int d^{2}\mathbf{r_{e}} d^{2}\mathbf{r_{h}} d^{2}\mathbf{r_{e'}} d^{2}\mathbf{r_{h'}} \Psi^{DX}_{\mathbf{Q}}(\mathbf{r_{e}},\mathbf{r_{h}})^{\ast}\Psi^{IX}_{\mathbf{Q'}}(\mathbf{r_{e'}},\mathbf{r_{h'}})^{\ast}V_{I}(\mathbf{r_{e}},\mathbf{r_{h}},\mathbf{r_{e'}},\mathbf{r_{h'}})\Psi^{DX}_{\mathbf{Q+q}}(\mathbf{r_{e'}},\mathbf{r_{h'}})\Psi^{IX}_{\mathbf{Q'-q}}(\mathbf{r_{e}},\mathbf{r_{h}}),\\
\label{Vexch_e}
V_{DX-IX}^{e-exch}&(\mathbf{Q},\mathbf{Q'},\mathbf{q})=\notag \\
&=  -\int d^{2}\mathbf{r_{e}} d^{2}\mathbf{r_{h}} d^{2}\mathbf{r_{e'}} d^{2}\mathbf{r_{h'}} \Psi^{DX}_{\mathbf{Q}}(\mathbf{r_{e}},\mathbf{r_{h}})^{\ast}\Psi^{IX}_{\mathbf{Q'}}(\mathbf{r_{e'}},\mathbf{r_{h'}})^{\ast}V_{I}(\mathbf{r_{e}},\mathbf{r_{h}},\mathbf{r_{e'}},\mathbf{r_{h'}})\Psi^{DX}_{\mathbf{Q+q}}(\mathbf{r_{e'}},\mathbf{r_{h}})\Psi^{IX}_{\mathbf{Q'-q}}(\mathbf{r_{e}},\mathbf{r_{h'}}),\\
\label{Vexch_h}
V_{DX-IX}^{h-exch}&(\mathbf{Q},\mathbf{Q'},\mathbf{q})= \notag \\
&= -\int d^{2}\mathbf{r_{e}} d^{2}\mathbf{r_{h}} d^{2}\mathbf{r_{e'}} d^{2}\mathbf{r_{h'}} \Psi^{DX}_{\mathbf{Q}}(\mathbf{r_{e}},\mathbf{r_{h}})^{\ast}\Psi^{IX}_{\mathbf{Q'}}(\mathbf{r_{e'}},\mathbf{r_{h'}})^{\ast}V_{I}(\mathbf{r_{e}},\mathbf{r_{h}},\mathbf{r_{e'}},\mathbf{r_{h'}})\Psi^{DX}_{\mathbf{Q+q}}(\mathbf{r_{e}},\mathbf{r_{h'}})\Psi^{IX}_{\mathbf{Q'-q}}(\mathbf{r_{e'}},\mathbf{r_{h}}).
\end{align}

\end{widetext}
Here $V_I$ is the Coulomb interaction between one direct exciton $(e,h)$ and one indirect exciton $(e',h')$. It can be expressed as
\begin{align*}
V_{I}(\mathbf{r_{e}},\mathbf{r_{h}},\mathbf{r_{e'}},\mathbf{r_{h'}})=\frac{e^2}{4\pi\epsilon\epsilon_{0}}\Big[\frac{1}{\sqrt{(\mathbf{r_{e}}-\mathbf{r_{e'}})^2 + L^2}}+\\+\frac{1}{|\mathbf{r_{h}}-\mathbf{r_{h'}}|}-\frac{1}{|\mathbf{r_{e}}-\mathbf{r_{h'}}|}-\frac{1}{\sqrt{(\mathbf{r_{h}}-\mathbf{r_{e'}})^{2}+L^{2}}}\Big],
\end{align*}
where $L$ is a separation distance between centers of coupled QWs and we used the narrow QW approximation.

Going to the center of mass coordinates of the excitons we set $\mathbf{R}=\beta_{e}\mathbf{r}^{e}+\beta_{h}\mathbf{r}^{h}$, and $\mathbf{r}=\mathbf{r}^{e}-\mathbf{r}^{h}$, where $\beta_{e}=m_{e}/(m_{e}+m_{h})$, $\beta_{h}=m_{h}/(m_{e}+m_{h})$. Separating the center of mass motion and the relative motion, we can write
\begin{equation}
\Psi^{DX/IX}_{\mathbf{Q}}(\mathbf{r_{e}}, \mathbf{r_{h}})=\frac{1}{\sqrt{A}} e^{i\mathbf{Q}\cdot \mathbf{R}} \phi_{DX/IX}(\mathbf{r}),
\end{equation}
where $\mathbf{Q}$ corresponds to the center-of-mass momentum of the exciton. The relative motion part of wave function $\phi(r)$ can be written using the variational procedure, where trial functions can be represented in several different forms, for example \cite{de-Leon_2000}
\begin{align}
\label{phi1}
&\phi_{1}(\mathbf{r})=\sqrt{\frac{2}{\pi}}\frac{1}{a}\exp\Big(-\frac{|\mathbf{r}|}{a}\Big),\\
\label{phi2}
&\phi_{2}(\mathbf{r})=\frac{1}{\sqrt{2\pi b(b+l)}}\exp\Big(-\frac{\sqrt{r^{2}+l^{2}}-l}{2b}\Big).
\end{align}
Here $a$ and $2b$ are quantities associated with indirect exciton Bohr radii and $l$ is a variational parameter reminiscent of the separation distance between the quantum wells.

\emph{Direct DX-IX interaction.}
First we calculate the matrix element for the direct interaction. We choose the second trial function, setting
\begin{equation}
\label{phiDX}
\phi_{DX}(\mathbf{r})=\frac{1}{\sqrt{2\pi b_{D}(b_{D}+l_{D})}}\exp\Big(-\frac{\sqrt{r^{2}+l_{D}^{2}}-l_{D}}{2b_{D}}\Big),
\end{equation}
\begin{equation}
\label{phiIX}
\phi_{IX}(\mathbf{r})=\frac{1}{\sqrt{2\pi b_{I}(b_{I}+l_{I})}}\exp\Big(-\frac{\sqrt{r^{2}+l_{I}^{2}}-l_{I}}{2b_{I}}\Big).
\end{equation}
Dropping the index $DX-IX$ for the sake of convenience, the integral (\ref{Vdir}) reads
\begin{widetext}
\begin{align}
\notag
&V_{dir}(\mathbf{Q},\mathbf{Q'},\mathbf{q})=\frac{1}{S^{2}}\frac{\exp(l_D/b_D)}{2\pi b_D(b_D+l_D)}\frac{\exp(l_I/b_I)}{2\pi b_I(b_I+l_I)}\int d^{2}\mathbf{r}_{e}d^{2}\mathbf{r}_{h}d^{2}\mathbf{r}_{e'}d^{2}\mathbf{r}_{h'}V_I(\mathbf{r_{e}},\mathbf{r_{h}},\mathbf{r_{e'}},\mathbf{r_{h'}}) \\
\notag &\exp[-i\mathbf{Q}(\beta_{e}\mathbf{r}_{e}+\beta_{h}\mathbf{r}_{h})-i\mathbf{Q'}(\beta_{e}\mathbf{r}_{e'}+\beta_{h}\mathbf{r}_{h'})+i(\mathbf{Q}+\mathbf{q})(\beta_{e}\mathbf{r}_{e}+\beta_{h}\mathbf{r}_{h})+i(\mathbf{Q'}-\mathbf{q})(\beta_{e}\mathbf{r}_{e'}+\beta_{h}\mathbf{r}_{h'})] \\ 
&\exp\Big(-\frac{\sqrt{(\mathbf{r}_{e}-\mathbf{r}_{h})^2+l_D^{2}}}{b_D}\Big)\exp\Big(-\frac{\sqrt{(\mathbf{r}_{e'}-\mathbf{r}_{h'})^2+l_I^{2}}}{b_I}\Big).
\label{Vdir_1}
\end{align}
\end{widetext}
Rewriting the integral using center-of-mass coordinates for both excitons, with $\mathbf{R}= \beta_{e}\mathbf{r}_{e}+ \beta_{h}\mathbf{r}_{h}$, $\mathbf{R'}= \beta_{e}\mathbf{r}_{e'}+ \beta_{h}\mathbf{r}_{h'}$, $\mathbf{\boldsymbol r}=\mathbf{r}_{e}-\mathbf{r}_{h}$ and $\mathbf{\boldsymbol r'}=\mathbf{r}_{e'}-\mathbf{r}_{h'}$, and defining the constant $C=\frac{e^{2}}{4\pi\epsilon\epsilon_{0}A}\frac{\exp(l_D/b_D)}{2\pi b_D(b_D+l_D)}\frac{\exp(l_I/b_I)}{2\pi b_I(b_I+l_I)}$, we get
\begin{widetext}
\begin{align}
\notag &V_{dir}(\mathbf{q})=\frac C S \int d^{2}\mathbf{\boldsymbol r}d^{2}\mathbf{\boldsymbol r'}d^{2}\mathbf{R}d^{2}\mathbf{R'} \exp\Big(-\frac{\sqrt{\mathbf{\boldsymbol r}^{2}+l_D^{2}}}{b_D}\Big) \exp\Big(-\frac{\sqrt{\mathbf{\boldsymbol r'}^{2}+l_I^{2}}}{b_I}\Big)\exp[i\mathbf{q}\cdot(\mathbf{R}-\mathbf{R'})] \\ \notag  
&\Big[\frac{1}{\sqrt{(\beta_{h}(\mathbf{\boldsymbol r}-\mathbf{\boldsymbol r'})+\mathbf{R}-\mathbf{R'})^2+L^2}}+\frac{1}{|-\beta_{e}(\mathbf{\boldsymbol r}-\mathbf{\boldsymbol r'})+\mathbf{R}-\mathbf{R'}|}-\frac{1}{|\beta_{h}\mathbf{\boldsymbol r}+\beta_{e}\mathbf{\boldsymbol r'}+\mathbf{R}-\mathbf{R'}|}- \\ &-\frac{1}{\sqrt{(-\beta_{e}\mathbf{\boldsymbol r}-\beta_{h}\mathbf{\boldsymbol r'}+\mathbf{R}-\mathbf{R'})^{2}+L^{2}}}\Big],
\label{Vdir_2}
\end{align}
\end{widetext}
where one can note that complex exponents with $\mathbf{Q}$ and $\mathbf{Q'}$ cancel each other. It is convenient to use the substitutions $\boldsymbol\xi=\mathbf{R}-\mathbf{R'}$, $\boldsymbol\sigma=(\mathbf{R}+\mathbf{R'})/2$. Then the integration over $\boldsymbol\sigma$ yields the area $S$ of the system. The integral (\ref{Vdir_2}) rewritten in a new variables reads
\begin{widetext}
\begin{align}
\notag
V_{dir}(\mathbf{q})&=C\int d^{2}\mathbf{\boldsymbol\xi}d^{2}\mathbf{\boldsymbol r}d^{2}\mathbf{\boldsymbol r'} \exp\Big(-\frac{\sqrt{\mathbf{\boldsymbol r}^{2}+l_D^{2}}}{b_D}\Big) \exp\Big(-\frac{\sqrt{\mathbf{\boldsymbol r'}^{2}+l_I^{2}}}{b_I}\Big)\exp[i\mathbf{q}\cdot \mathbf{\boldsymbol\xi}] \\ 
&\Big[\frac{1}{\sqrt{(\mathbf{\boldsymbol\xi} + \beta_{h}(\mathbf{\boldsymbol r}-\mathbf{\boldsymbol r'}))^2+L^2}}+\frac{1}{|\mathbf{\boldsymbol\xi}-\beta_{e}(\mathbf{\boldsymbol r}-\mathbf{\boldsymbol r'})|}-\frac{1}{|\mathbf{\boldsymbol\xi}+\beta_{h}\mathbf{\boldsymbol r}+\beta_{e}\mathbf{\boldsymbol r'}|}-\frac{1}{\sqrt{(\mathbf{\boldsymbol\xi}-\beta_{e}\mathbf{\boldsymbol r}-\beta_{h}\mathbf{\boldsymbol r'})^{2}+L^{2}}}\Big].
\label{Vdir_3}
\end{align}
\end{widetext}
The expression (\ref{Vdir_3}) represents a sum of four integrals for electron-electron, hole-hole and electron-hole mixed interaction,
\begin{equation}
V_{dir}=\Big[\mathcal{I}_{ee'}+\mathcal{I}_{hh'}+\mathcal{I}_{eh'}+\mathcal{I}_{he'} \Big].
\label{Vdir_4}
\end{equation}
These integrals can be calculated separately using the standard integrals
\begin{align*}
&\int_{0}^{2\pi} e^{i x \sin \theta}d\theta = 2\pi J_0 (x), \\
&\int_0^\infty \frac{x}{\sqrt{a^2 +x^2}} J_0(bx) dx = \frac{e^{-ab}}{b}.
\end{align*}
We start with $\mathcal{I}_{ee'}$, and perform the change of variables $\boldsymbol \chi = \boldsymbol\xi +\beta_h (\mathbf{\boldsymbol{r-r'}})$:
\begin{widetext}
\begin{align}
\notag
\mathcal{I}_{ee'}(\mathbf{q})&=C\int d^{2}\mathbf{\boldsymbol r}d^{2}\mathbf{\boldsymbol r'}d^{2}\mathbf{\boldsymbol\xi}
\exp\Big(-\frac{\sqrt{\mathbf{\boldsymbol r}^{2}+l_D^{2}}}{b_D}\Big) \exp\Big(-\frac{\sqrt{\mathbf{\boldsymbol\ r'}^{2}+l_I^{2}}}{b_I}\Big)\exp[i\mathbf{q} \cdot\boldsymbol\xi]\Big[\frac{1}{\sqrt{(\boldsymbol\xi + \beta_{h}(\mathbf{\boldsymbol\ r}-\mathbf{\boldsymbol\ r'}))^{2}+L^{2}}}\Big] \\
\notag
&=C\int d^{2}\mathbf{\boldsymbol r}d^{2}\mathbf{\boldsymbol r'}d^{2}\mathbf{\boldsymbol\chi}
\exp\Big(-\frac{\sqrt{\mathbf{\boldsymbol r}^{2}+l_D^{2}}}{b_D}\Big) \exp\Big(-\frac{\sqrt{\mathbf{\boldsymbol\ r'}^{2}+l_I^{2}}}{b_I}\Big)\exp[i\mathbf{q} \cdot\boldsymbol\chi -i\beta_h \mathbf{q} \cdot(\mathbf{\boldsymbol{r-r'}})]\frac{1}{\sqrt{\mathbf{\boldsymbol\chi}^{2}+L^{2}}} \\
\notag
&=C\int d^{2}\mathbf{\boldsymbol r}d^{2}\mathbf{\boldsymbol r'}
\exp\Big(-\frac{\sqrt{\mathbf{\boldsymbol r}^{2}+l_D^{2}}}{b_D}-\frac{\sqrt{\mathbf{\boldsymbol\ r'}^{2}+l_I^{2}}}{b_I}\Big)\exp[-i\beta_h \mathbf{q} \cdot(\mathbf{\boldsymbol{r-r'}})] \int_0^\infty d\chi \frac{\chi}{\sqrt{\chi^{2}+L^{2}}}\int_0^{2\pi}d\theta e^{i q\chi \cos{\theta}} \\
\notag
&=2\pi C\int d^{2}\mathbf{\boldsymbol r}d^{2}\mathbf{\boldsymbol r'}
\exp\Big(-\frac{\sqrt{\mathbf{\boldsymbol r}^{2}+l_D^{2}}}{b_D}-\frac{\sqrt{\mathbf{\boldsymbol\ r'}^{2}+l_I^{2}}}{b_I}\Big)\exp[-i\beta_h \mathbf{q} \cdot(\mathbf{\boldsymbol{r-r'}})] \int_0^\infty d\chi \frac{\chi}{\sqrt{\chi^{2}+L^{2}}}J_0 (q\chi) \\
\notag
&=2\pi C \frac {\exp(-qL)} q \int d^{2}\mathbf{\boldsymbol r}
\exp\Big(-\frac{\sqrt{\mathbf{\boldsymbol r}^{2}+l_D^{2}}}{b_D}\Big)\exp[-i\beta_h \mathbf{q} \cdot\mathbf{\boldsymbol{r}}]\int d^{2}\mathbf{\boldsymbol r'} \exp\Big(-\frac{\sqrt{\mathbf{\boldsymbol\ r'}^{2}+l_I^{2}}}{b_I}\Big)\exp[i\beta_h \mathbf{q} \cdot \mathbf{\boldsymbol r'}]\\
&=2\pi C \frac {\exp(-qL)} q I_D(\mathbf{q}) I_I(\mathbf{q}).
\end{align}
\end{widetext}
The integral is decomposed into a product of two integrals, which are evaluated as

\begin{align}
\notag I&_{D/I}(\mathbf{q})= \\
\notag  &=\int_0^\infty d r \cdot r\exp\Big(-\frac{\sqrt{ r^{2}+l_{D/I}^{2}}}{b_{D/I}}\Big)\int_{0}^{2\pi}d\theta e^{\mp i\beta_{h}q r \cos\theta} \\
&= 2\pi\int_0^\infty d r \cdot r\exp\Big(-\frac{\sqrt{ r^{2}+l_{D/I}^{2}}}{b_{D/I}}\Big)J_{0}(\beta_{h}q r).
\label{Ir}
\end{align}
It is not possible to calculate integral (\ref{Ir}) analytically in the general case, but we are interested in $q \rightarrow 0$ limit. In this limit the zeroeth order Bessel function equals one. We make the change of variables $x = \sqrt{r^{2}+r_{0}^2}$ and integrate by parts:
\begin{align*}
I_{D/I}(\mathbf{q}\rightarrow 0)&=2\pi\int_{0}^{+\infty} d r\cdot r\exp\Big(-\frac{\sqrt{ r^{2}+l_{D/I}^{2}}}{b_{D/I}}\Big)\\
&=2\pi\int_{l_{D/I}}^{+\infty} dx\cdot x \exp\Big(-\frac x {b_{D/I}}\Big) \\
&= 2\pi \exp\Big(-\frac{l_{D/I}}{b_{D/I}}\Big)b_{D/I}(b_{D/I}+l_{D/I}).
\end{align*}

Finally, the integral $\mathcal{I}_{ee'}$ reads
\begin{widetext}
\begin{align}
\notag \mathcal{I}_{ee'}^{\mathbf{q}\rightarrow 0} &= 2\pi\frac {e^{-qL}} q \frac{e^{2}}{4\pi\epsilon\epsilon_{0}S}\frac{\exp(l_D/b_D)}{2\pi b_D(b_D+l_D)}\frac{\exp(l_I/b_I)}{2\pi b_I(b_I+l_I)} (2\pi)^2  \exp\Big(-\frac{l_{I}}{b_{I}}-\frac{l_{D}}{b_{D}}\Big) b_{D}(b_{D}+l_{D}) b_{I}(b_{I}+l_{I}) \\
&= \frac{e^{2}}{2\epsilon\epsilon_{0}S}\frac {e^{-qL}} q.
\end{align}
\end{widetext}
The three remaining integrals can be calculated in the same way, using different change of variables. For the integral $\mathcal{I}_{hh'}$ we employ the substitution $\boldsymbol \chi = \boldsymbol\xi -\beta_e (\mathbf{\boldsymbol{r-r'}})$, yielding the result
\begin{equation*}
\mathcal{I}_{hh'}^{\mathbf{q}\rightarrow 0}= \frac{e^{2}}{2\epsilon\epsilon_{0}S}\frac {1} q.
\end{equation*}
The integral $\mathcal{I}_{eh'}$, using $\boldsymbol{\chi} = \boldsymbol{\xi} + \beta_h \mathbf{\boldsymbol{r}} + \beta_e\mathbf{\boldsymbol{r'}} $, results in
\begin{equation*}
\mathcal{I}_{eh'}^{\mathbf{q}\rightarrow 0}=- \frac{e^{2}}{2\epsilon\epsilon_{0}S}\frac {1} q ,
\end{equation*}
and $\mathcal{I}_{he'}$ calculated with $\boldsymbol{\chi} = \boldsymbol{\xi} - \beta_e \mathbf{\boldsymbol{r}} - \beta_h \mathbf{\boldsymbol{r'}}$ gives
\begin{equation*}
\mathcal{I}_{he'}^{\mathbf{q}\rightarrow 0}= -\frac{e^{2}}{2\epsilon\epsilon_{0}S}\frac {e^{-qL}} q .
\end{equation*}
Finally, the sum of the four integrals is
\begin{eqnarray}
V^{q\rightarrow 0}_{dir}=\frac{e^2}{2\epsilon\epsilon_{0}S}\Big[\frac {e^{-qL}} q +\frac {1} q - \frac {1} q- \frac {e^{-qL}} q \Big] = 0.
\label{Vdir_fin}
\end{eqnarray}
This corresponds to the fact that \emph{direct interaction between direct and indirect exciton vanishes} in $q \rightarrow 0$ limit, on the contrary to IX-IX case, where dipole-dipole interaction gives the main contribution.\cite{IXIXint, de-Leon_2001}

\emph{Exciton exchange DX-IX interaction.} The interaction matrix element due to simultaneous exchange of electron and hole between two excitons, which is referred as exciton exchange $V_{X}^{exch}$, can be calculated using same trial functions as for the direct matrix element. It reads

\begin{widetext}
\begin{align}
\notag &V_{X}^{exch}(\mathbf{Q},\mathbf{Q'},\mathbf{q})=\frac{1}{S^{2}}\frac{\exp(l_D/b_D)}{2\pi b_D(b_D+l_D)}\frac{\exp(l_I/b_I)}{2\pi b_I(b_I+l_I)}\int d^{2}\mathbf{r}_{e}d^{2}\mathbf{r}_{h}d^{2}\mathbf{r}_{e'}d^{2}\mathbf{r}_{h'}V_I(\mathbf{r_{e}},\mathbf{r_{h}},\mathbf{r_{e'}},\mathbf{r_{h'}}) \\
\notag &\exp[-i\mathbf{Q}(\beta_{e}\mathbf{r}_{e}+\beta_{h}\mathbf{r}_{h})-i\mathbf{Q'}(\beta_{e}\mathbf{r}_{e'}+\beta_{h}\mathbf{r}_{h'})+i(\mathbf{Q}+\mathbf{q})(\beta_{e}\mathbf{r}_{e'}+\beta_{h}\mathbf{r}_{h'})+i(\mathbf{Q'}-\mathbf{q})(\beta_{e}\mathbf{r}_{e}+\beta_{h}\mathbf{r}_{h})]\\
\notag &\exp\Big(-\frac{\sqrt{(\mathbf{r}_{e}-\mathbf{r}_{h})^2+l_D^{2}}}{b_D}\Big)\exp\Big(-\frac{\sqrt{(\mathbf{r}_{e'}-\mathbf{r}_{h'})^2+l_I^{2}}}{b_I}\Big)\exp\Big(-\frac{\sqrt{(\mathbf{r}_{e'}-\mathbf{r}_{h'})^2+l_D^{2}}}{b_D}\Big)\exp\Big(-\frac{\sqrt{(\mathbf{r}_{e}-\mathbf{r}_{h})^2+l_I^{2}}}{b_I}\Big)\\
\notag &=C\int d^{2}\mathbf{\boldsymbol\xi}d^{2}\mathbf{\boldsymbol r}d^{2}\mathbf{\boldsymbol r'} \exp\Big(-\frac{\sqrt{\mathbf{\boldsymbol r}^{2}+l_D^{2}}}{b_D}-\frac{\sqrt{\mathbf{\boldsymbol r'}^{2}+l_I^{2}}}{b_I}-\frac{\sqrt{\mathbf{\boldsymbol r}^{2}+l_I^{2}}}{b_I}-\frac{\sqrt{\mathbf{\boldsymbol r'}^{2}+l_D^{2}}}{b_D}\Big)\exp[-i(\mathbf{q}+\Delta\mathbf{Q}) \cdot \mathbf{\boldsymbol\xi}] \\
 &\Big[\frac{1}{\sqrt{(\mathbf{\boldsymbol\xi} + \beta_{h}(\mathbf{\boldsymbol r}-\mathbf{\boldsymbol r'}))^2+L^2}}+\frac{1}{|\mathbf{\boldsymbol\xi}-\beta_{e}(\mathbf{\boldsymbol r}-\mathbf{\boldsymbol r'})|}-\frac{1}{|\mathbf{\boldsymbol\xi}+\beta_{h}\mathbf{\boldsymbol r}+\beta_{e}\mathbf{\boldsymbol r'}|}-\frac{1}{\sqrt{(\mathbf{\boldsymbol\xi}-\beta_{e}\mathbf{\boldsymbol r}-\beta_{h}\mathbf{\boldsymbol r'})^{2}+L^{2}}}\Big],
\end{align}
\end{widetext}
where $C=\frac{e^{2}}{4\pi\epsilon\epsilon_{0}S}\frac{\exp(l_D/b_D)}{2\pi b_D(b_D+l_D)}\frac{\exp(l_I/b_I)}{2\pi b_I(b_I+l_I)}$ and $\boldsymbol\xi=\mathbf{R}-\mathbf{R'}$. Additionally, we defined exchanged momentum between excitons as $\Delta\mathbf{Q}=\mathbf{Q}-\mathbf{Q'}$. We are interested in small values of $\Delta\mathbf{Q}$, taking the limit as it approaches zero. It is therefore convenient to define momentum $\mathbf{K} = \Delta\mathbf{Q}+\mathbf{q}$, and study its long wavelength limit.

We proceed in the similar way as with the direct interaction, separating the integral into four parts for different interaction terms:
\begin{equation*}
V_X^{exch}(\mathbf{K})=\Big[\mathcal{I}_{ee'}^{'}+\mathcal{I}_{hh'}^{'}+\mathcal{I}_{eh'}^{'}+\mathcal{I}_{he'}^{'} \Big].
\label{Vexch_partition}
\end{equation*}
We start by calculating $\mathcal{I}_{ee'}^{'}$, using the same change of variables $\boldsymbol \chi = \boldsymbol\xi +\beta_h (\mathbf{\boldsymbol{r-r'}})$ as before:
\begin{widetext}
\begin{align}
\notag \mathcal{I}_{ee'}^{'}=&C\int d^{2}\mathbf{\boldsymbol\xi}d^{2}\mathbf{\boldsymbol r}d^{2}\mathbf{\boldsymbol r'} \exp\Big(-\frac{\sqrt{\mathbf{\boldsymbol r}^{2}+l_D^{2}}}{b_D}-\frac{\sqrt{\mathbf{\boldsymbol r'}^{2}+l_I^{2}}}{b_I}-\frac{\sqrt{\mathbf{\boldsymbol r}^{2}+l_I^{2}}}{b_I}-\frac{\sqrt{\mathbf{\boldsymbol r'}^{2}+l_D^{2}}}{b_D}\Big)e^{-i\mathbf{K} \cdot \mathbf{\boldsymbol\xi}}\frac{1}{\sqrt{(\mathbf{\boldsymbol\xi} + \beta_{h}(\mathbf{\boldsymbol r}-\mathbf{\boldsymbol r'}))^2+L^2}} \\
\notag =&C\int d^{2}\mathbf{\boldsymbol\chi}d^{2}\mathbf{\boldsymbol r}d^{2}\mathbf{\boldsymbol r'} \exp\Big(-\frac{\sqrt{\mathbf{\boldsymbol r}^{2}+l_D^{2}}}{b_D}-\frac{\sqrt{\mathbf{\boldsymbol r'}^{2}+l_I^{2}}}{b_I}-\frac{\sqrt{\mathbf{\boldsymbol r}^{2}+l_I^{2}}}{b_I}-\frac{\sqrt{\mathbf{\boldsymbol r'}^{2}+l_D^{2}}}{b_D}\Big) \\
\notag & \exp[-i\mathbf{K} \cdot \mathbf{\boldsymbol\chi}] \exp[i\beta_h \mathbf{K}\cdot (\mathbf{\boldsymbol r}-\mathbf{\boldsymbol r'})]
 \frac{1}{\sqrt{\boldsymbol\chi^2+L^2}}  \\
\notag =&C\int d^{2}\mathbf{\boldsymbol r}d^{2}\mathbf{\boldsymbol r'} \exp\Big(-\frac{\sqrt{\mathbf{\boldsymbol r}^{2}+l_D^{2}}}{b_D}-\frac{\sqrt{\mathbf{\boldsymbol r'}^{2}+l_I^{2}}}{b_I}-\frac{\sqrt{\mathbf{\boldsymbol r}^{2}+l_I^{2}}}{b_I}-\frac{\sqrt{\mathbf{\boldsymbol r'}^{2}+l_D^{2}}}{b_D}\Big) \exp[i\beta_h \mathbf{K}\cdot (\mathbf{\boldsymbol r}-\mathbf{\boldsymbol r'})] \\
\notag & \int_0^\infty d\chi \frac{\chi}{\sqrt{\chi^2+L^2}} \int_0^{2\pi}d\theta\exp(-i\beta_h K\chi\cos\theta)  \\
\notag =&C\int d^{2}\mathbf{\boldsymbol r}d^{2}\mathbf{\boldsymbol r'} \exp\Big(-\frac{\sqrt{\mathbf{\boldsymbol r}^{2}+l_D^{2}}}{b_D}-\frac{\sqrt{\mathbf{\boldsymbol r'}^{2}+l_I^{2}}}{b_I}-\frac{\sqrt{\mathbf{\boldsymbol r}^{2}+l_I^{2}}}{b_I}-\frac{\sqrt{\mathbf{\boldsymbol r'}^{2}+l_D^{2}}}{b_D}\Big) \exp[i\beta_h \mathbf{K}\cdot (\mathbf{\boldsymbol r}-\mathbf{\boldsymbol r'})] \int_0^\infty d\chi\frac{\chi J_0(\beta_h K\chi)}{\sqrt{\chi^2+L^2}}\\ 
\notag =&2\pi C\frac {e^{-KL}} K\int d^{2}\mathbf{\boldsymbol r}e^{i\beta_h \mathbf{K}\cdot \mathbf{\boldsymbol r}} \exp\Big(-\frac{\sqrt{\mathbf{\boldsymbol r}^{2}+l_D^{2}}}{b_D}-\frac{\sqrt{\mathbf{\boldsymbol r}^{2}+l_I^{2}}}{b_I}\Big)\int d^{2}\mathbf{\boldsymbol r'}e^{-i\beta_h \mathbf{K}\cdot \mathbf{\boldsymbol r'}}\exp\Big(-\frac{\sqrt{\mathbf{\boldsymbol r'}^{2}+l_D^{2}}}{b_D}-\frac{\sqrt{\mathbf{\boldsymbol r'}^{2}+l_I^{2}}}{b_I}\Big)\\
=& 2\pi C\frac {e^{-KL}} K I_+(\mathbf{K}) I_-(\mathbf{K}),
\end{align}
\end{widetext}
and we have written the total integral as a product of two integrals, which can be written as
\begin{widetext}
\begin{align*}
\notag
I_{+/-}(\mathbf{K})&=\int_0^\infty dr \: r\exp\Big(-\frac{\sqrt{ r^{2}+l_{D}^{2}}}{b_{D}}\Big)\exp\Big(-\frac{\sqrt{ r^{2}+l_{I}^{2}}}{b_{I}}\Big)\int_{0}^{2\pi}d\theta e^{\pm i\beta_{h}K r \cos\theta} \\
&=2\pi\int_0^\infty dr \: r\exp\Big(-\frac{\sqrt{ r^{2}+l_{D}^{2}}}{b_{D}}\Big)\exp\Big(-\frac{\sqrt{ r^{2}+l_{I}^{2}}}{b_{I}}\Big)J_{0}(\beta_{h}K r) \\
&\rightarrow 2\pi\int_0^\infty dr\: r\exp\Big(-\frac{\sqrt{ r^{2}+l_{D}^{2}}}{b_{D}}\Big)\exp\Big(-\frac{\sqrt{ r^{2}+l_{I}^{2}}}{b_{I}}\Big) := I_X
\label{I_pm}
\end{align*}
\end{widetext}
in the limit $K \rightarrow 0$. As $\mathbf{K} = \mathbf{q} + \Delta\mathbf{Q}$ we can write
\[ \mathcal{I}_{ee'}^{'}(\mathbf{q}\rightarrow 0,\Delta\mathbf{Q}\rightarrow 0) = 2\pi C I_X^2 \frac {e^{-KL}} K .\]
Repeating the procedure for other three terms we find
\begin{align*}
\mathcal{I}_{hh'}^{'}(\mathbf{q}\rightarrow 0,\Delta\mathbf{Q}\rightarrow 0) &= 2\pi C I_X^2 \frac {1} K ,\\
\mathcal{I}_{eh'}^{'}(\mathbf{q}\rightarrow 0,\Delta\mathbf{Q}\rightarrow 0) &=-2\pi C I_X^2 \frac {1} K ,\\
\mathcal{I}_{he'}^{'}(\mathbf{q}\rightarrow 0,\Delta\mathbf{Q}\rightarrow 0) &=-2\pi C I_X^2 \frac {e^{-KL}} K .
\end{align*}
The total exciton exchange term becomes
\[ V_X(K\rightarrow 0) = 2\pi C I_X^2\Big[\frac {e^{-KL}} K +\frac {1} K - \frac {1} K- \frac {e^{-KL}} K \Big] = 0. \]

\emph{Electron and hole exchange DX-IX interaction.}
The general form of the electron exchange interaction between direct and indirect exciton is given by Eq. (\ref{Vexch_e}).
For the calculations it is more convenient to choose the indirect exciton wave function written in the form (\ref{phi1}). Thus the exchange interaction matrix element reads
\begin{widetext}
\begin{align}
\notag V_{e}^{exch}(\mathbf{Q},\mathbf{Q'},\mathbf{q})=&-\frac{e^2}{4\pi\epsilon\epsilon_{0}} \frac 1 {S^2}\frac{4}{\pi^2} \frac{1}{a_{DX}^2}\frac{1}{a_{IX}^2}\int d^{2}\mathbf{r_{e}}d^{2}\mathbf{r_{h}}d^{2}\mathbf{r_{e'}}d^{2}\mathbf{r_{h'}} \\
\notag &\cdot \exp\Big(-\frac{|\mathbf{r}_{e}-\mathbf{r}_{h}|+ |\mathbf{r}_{e'}-\mathbf{r}_{h}|}{a_{DX}}-\frac{|\mathbf{r}_{e'}-\mathbf{r}_{h'}|+|\mathbf{r}_{e}-\mathbf{r}_{h'}|}{a_{IX}}\Big)\\  
\notag &\cdot \exp[-i\mathbf{Q}(\beta_{e}\mathbf{r}_{e}+\beta_{h}\mathbf{r}_{h})-i\mathbf{Q'}(\beta_{e}\mathbf{r}_{e'}+\beta_{h}\mathbf{r}_{h'})+i(\mathbf{Q}+\mathbf{q})(\beta_{e}\mathbf{r}_{e}+\beta_{h}\mathbf{r}_{h})+i(\mathbf{Q'}-\mathbf{q})(\beta_{e}\mathbf{r}_{e'}+\beta_{h}\mathbf{r}_{h'})] \\
&\cdot \Big[\frac{1}{\sqrt{(\mathbf{r_{e}}-\mathbf{r_{e'}})^2 + L^2}}+\frac{1}{|\mathbf{r_{h}}-\mathbf{r_{h'}}|}-\frac{1}{|\mathbf{r_{e}}-\mathbf{r_{h'}}|}-\frac{1}{\sqrt{(\mathbf{r_{h}}-\mathbf{r_{e'}})^{2}+L^{2}}}\Big].
\end{align}
\end{widetext}

The exact calculation of exchange integral is straightforward but tedious. Starting with the same substitutions as for direct interaction calculation, and defining new variables 
\begin{align*}
\mathbf{y}_{1}&=(\boldsymbol\xi-\beta_{e}\boldsymbol r-\beta_{h}\boldsymbol r')/a_{DX},\\
\mathbf{y}_{2}&=(\boldsymbol\xi+\beta_{h}\boldsymbol r+\beta_{e}\boldsymbol r')/a_{DX},\\
\mathbf{x}   &=\boldsymbol r/a_{DX},
\end{align*}
one gets the final expression of electron exchange interaction
\begin{equation}
V_{e}^{exch}=-\frac{e^{2}}{\pi^3 \epsilon\epsilon_{0}S}\frac{a_{DX}^3}{a_{IX}^2}\mathcal{I}_{e}^{exch}(\Delta Q,q,\Theta,\beta_{e},a_{DX},\alpha).
\label{Vexch_e_fin}
\end{equation}
Here $\alpha= a_{DX}/a_{IX}$ is the ratio between the direct and indirect exciton Bohr radii, and $\Theta$ is the angle between $\Delta\mathbf{Q}$ and $\mathbf{q}$. The integral is given by
\begin{widetext}
\begin{align}\notag
&\mathcal{I}_{e}^{exch}(\Delta Q,q,\Theta,\beta_{e},a_{DX},\alpha)=\int_{0}^{\infty}dx\int_{0}^{2\pi}d\Theta_{x}\int_{0}^{\infty}dy_{1}\int_{0}^{2\pi}d\Theta_{1}\int_{0}^{\infty}dy_{2}\int_{0}^{2\pi}d\Theta_{2} xy_{1}y_{2} \\ \notag
& \cos\Big[a_{DX}\beta_{e}\Delta Q[x\cos(\Theta-\Theta_{x})+y_{1}\cos(\Theta-\Theta_{1})]+a_{DX}q[x\cos\Theta_{x}+\beta_{e}y_{1}\cos\Theta_{1}-(1-\beta_{e})y_{2}\cos\Theta_{2}]\Big]  \\ \notag
& \exp(-x-y_1-\alpha y_{2}) \exp\Big[-\alpha\sqrt{(y_{2}\cos\Theta_{2}-y_{1}\cos\Theta_{1}-x\cos\Theta_{x})^{2}+(y_{2}\sin\Theta_{2}-y_{1}\sin\Theta_{1}-x\sin\Theta_{x})^{2}}\Big] \\
&\Big[\frac{1}{\sqrt{y_{1}^{2}+x^{2}+2y_{1}x\cos(\Theta_{1}-\Theta_{x}) +\widetilde{L}^2}}+\frac{1}{\sqrt{y_{2}^{2}+x^{2}-2y_{2}x\cos(\Theta_{2}-\Theta_{x})}}-\frac{1}{y_{2}}-\frac{1}{\sqrt{y_{1}^{2}+\widetilde{L}^{2}}}\Big],
\label{Iexch}
\end{align}
\end{widetext}
where $\widetilde{L}= L/a_{DX}$. The calculation of exchange integral requires numerical integration with a multidimensional Monte Carlo algorithm.

The same procedure for the hole exchange term yields 
\begin{equation}
V_{h}^{exch}=-\frac{e^{2}}{\pi^3 \epsilon\epsilon_{0}S}\frac{a_{DX}^3}{a_{IX}^2}\mathcal{I}_{h}^{exch}(\Delta Q,q,\Theta,\beta_{e},a_{DX},\alpha),
\label{Vexch_h_fin}
\end{equation}
where one needs to evaluate numerically the expression
\begin{widetext}
\begin{align}\notag
&\mathcal{I}_{h}^{exch}(\Delta Q,q,\Theta,\beta_{e},a_{DX},\alpha)=\int_{0}^{\infty}dx\int_{0}^{2\pi}d\Theta_{x}\int_{0}^{\infty}dy_{1}\int_{0}^{2\pi}d\Theta_{1}\int_{0}^{\infty}dy_{2}\int_{0}^{2\pi}d\Theta_{2} xy_{1}y_{2} \\ \notag
& \cos\Big[a_{DX}(1-\beta_{e})\Delta Q[-x\cos(\Theta-\Theta_{x})+y_{2}\cos(\Theta-\Theta_{2})]+a_{DX}q[-x\cos\Theta_{x}-\beta_{e}y_{1}\cos\Theta_{1}+(1-\beta_{e})y_{2}\cos\Theta_{2}]\Big] \\ \notag
& \exp(-x-\alpha y_1- y_{2}) \exp\Big[-\alpha\sqrt{(y_{2}\cos\Theta_{2}-y_{1}\cos\Theta_{1}-x\cos\Theta_{x})^{2}+(y_{2}\sin\Theta_{2}-y_{1}\sin\Theta_{1}-x\sin\Theta_{x})^{2}}\Big] \\
& \Big[\frac{1}{\sqrt{y_{1}^{2}+x^{2}+2y_{1}x\cos(\Theta_{1}-\Theta_{x}) +\widetilde{L}^2}}+\frac{1}{\sqrt{y_{2}^{2}+x^{2}-2y_{2}x\cos(\Theta_{2}-\Theta_{x})}}-\frac{1}{y_{2}}-\frac{1}{\sqrt{y_{1}^{2}+\widetilde{L}^{2}}}\Big].
\label{Iexch-h}
\end{align}
\end{widetext}

\begin{figure}[h!]
\includegraphics[width=0.42\textwidth]{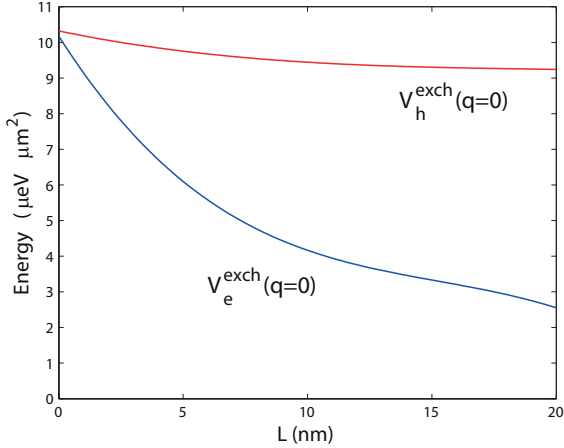}
\caption{(Color online) Electron and hole exchange terms of the DX-IX interaction constant plotted as a function of separation between centers of quantum wells. The hole exchange term is larger than the electron exchange term for all values of well separation $L$ calculated.}
\label{fig:matrix_elements}
\end{figure}

The total direct-indirect exciton interaction constant can be written as $V_{DI} = V_e^{exch} + V_h^{exch}$. The two non-zero terms are shown as a function of quantum well separation in Fig. \ref{fig:matrix_elements}.


\begin{thebibliography}{99}

\bibitem{Siegel} P. H. Siegel, IEEE Trans. Microw. Theory Techn. \textbf{50}, 910 (2002).

\bibitem{MilesNATO} R. E. Miles, P. Harrison, and D. Lippens (eds), \textit{Terahertz Sources and Systems} Vol. 27 (NATO Science Series II, Kluwer, Dordrecht, 2001).

\bibitem{spectroscopy} P. F. Taday, I. V. Bradley, D. D. Arnone, and M. Pepper, J. Pharm. Sci. \textbf{92}, 831 (2003).

\bibitem{Gunn} H. Eisele, A. Rydberg, and G. I. Haddad, IEEE Trans. Microw. Theory Techn. \textbf{48}, 626 (2000).

\bibitem{Gold} S. H. Gold and G. S. Nusinovich, Rev. Sci Instrum. \textbf{68}, 3945 (1997).

\bibitem{femtosecond} J. Shan and T. F. Heinz, Topics Appl. Phys. \textbf{92}, 59 (2004).

\bibitem{SovietQCL} R. F. Kazarinov and R. A. Suris, Sov. Phys. Semicond. \textbf{5}, 207 (1971).

\bibitem{ScienceQCL} J. Faist, F. Capasso, D. L. Sivco, C. Sirtori, A. L. Hutchinson, and A. Y. Cho, Science \textbf{264}, 553 (1994).

\bibitem{IEEEQCL} M. Razeghi, IEEE J. Sel. Top. Quantum Electron. \textbf{15}, 941 (2009).

\bibitem{crystal} Z. Jiang and X.-C. Zhang, IEEE Trans. Microw. Theory Techn. \textbf{47}, 2644 (1999).

\bibitem{KavokinBook} A. V. Kavokin, J. J. Baumberg, G. Malpuech, and F. P. Laussy, \textit{Microcavities} (Oxford University Press, Oxford, 2007).

\bibitem{PolaritonDevices} T. C. H. Liew, I. A. Shelykh, and G. Malpuech, Physica E \textbf{43}, 1543 (2011).

\bibitem{KVKavokin} K. V. Kavokin, M. A. Kaliteevski, R. A. Abram, A. V. Kavokin, S. Sharkova, and I. A. Shelykh, Appl. Phys. Lett. \textbf{97}, 201111 (2010).

\bibitem{bistableTransition2} E. del Valle and A. V. Kavokin, Phys. Rev. B \textbf{83}, 193303 (2011).

\bibitem{Savenko} I. G. Savenko, I. A. Shelykh, and M. A. Kaliteevski, Phys. Rev. Lett. \textbf{107}, 027401 (2011).

\bibitem{1S2P} A. V. Kavokin, I. A. Shelykh, T. Taylor, and M. M. Glazov, Phys. Rev. Lett. \textbf{108}, 197401 (2012).

\bibitem{bosonicQCL} T. C. H. Liew, M. M. Glazov, K. V. Kavokin, I. A. Shelykh, M. A. Kaliteevski, and A. V. Kavokin, Phys. Rev. Lett. \textbf{110}, 047402 (2013).

\bibitem{dipolaritonTHz} O. Kyriienko, A. V. Kavokin, and I. A. Shelykh, arXiv:1211.0688.

\bibitem{ChristmannAPL} G. Christmann, A. Askitopoulos, G. Deligeorgis, Z. Hatzopoulos, S. I. Tsintzos, P. G. Savvidis, and J. J. Baumberg, Appl. Phys. Lett. \textbf{98}, 081111 (2011).

\bibitem{Cristofolini} P. Cristofolini, G. Christmann, S. I. Tsintzos, G. Deligeorgis, G. Konstantinidis, Z. Hatzopoulos, P. G. Savvidis, and J. J. Baumberg, Science \textbf{336}, 704 (2012).

\bibitem{indirect1976} Yu. E. Lozovik and V. I. Yudson, Sov. Phys. JETP \textbf{44}, 389 (1976).

\bibitem{EtuningPhysRev} L. V. Butov, A. L. Ivanov, A. Imamoglu, P. B. Littlewood, A. A. Shashkin, V. T. Dolgopolov, K. L. Campman, and A. C. Gossard, Phys. Rev. Lett. \textbf{86}, 5608 (2001).

\bibitem{EtuningNature} A. A. High, J. R. Leonard, A. T. Hammack, M. M. Fogler, L. V. Butov, A. V. Kavokin, K. L. Campman, and A. C. Gossard, Nature \textbf{483}, 584 (2012).

\bibitem{Baas} A. Baas, J.-P. Karr, H. Eleuch, and E. Giacobino, Phys. Rev. A \textbf{69}, 023809 (2004).

\bibitem{Gippius1} N. A. Gippius, S. G. Tikhodeev, V. D. Kulakovskii, D. N. Krizhanovskii, and A. I. Tartakovskii, Europhys. Lett. \textbf{67}, 997 (2004).

\bibitem{Whittaker} D. M. Whittaker, Phys. Rev. B, \textbf{71}, 115301 (2005).

\bibitem{Gippius2} N. A. Gippius, I. A. Shelykh, D. D. Solnyshkov, S. S. Gavrilov, Y. G. Rubo, A. V. Kavokin, S. G. Tikhodeev, and G. Malpuech, Phys. Rev. Lett. \textbf{98}, 236401 (2007).


\bibitem{Paraiso} T. K. Paraiso, M. Wouters. Y. Leger, F. Morier-Genoud, and B. Deveaud-Pledran, Nature Mater. \textbf{9}, 655 (2010).

\bibitem{Shelykh} I. A. Shelykh, T. C. H. Liew, and A. V. Kavokin, Phys. Rev. Lett. \textbf{100}, 116401 (2008).

\bibitem{Schumacher} S. Schumacher, N. H. Kwong, R. Binder, and A. L. Smirl, Phys. Status Solidi RRL \textbf{3}, 10 (2009).

\bibitem{Adrados} C. Adrados, T. C. H. Liew, A. Amo, M. D. Martin, D. Sanvitto, C. Anton, E. Giacobino, A. Kavokin, A. Bramati, and L. Vina, Phys. Rev. Lett. \textbf{107}, 146402 (2011).

\bibitem{Giorgi} M. De Giorgi, D. Ballarini, E. Cancellieri, F. M. Marchetti, M. H. Szymanska, C. Tejedor, R. Cingolani, E. Giacobino, A. Bramati, G. Gigli, and D. Sanvitto, Phys. Rev. Lett. \textbf{109}, 266407 (2012).

\bibitem{Sarchi} D. Sarchi, I. Carusotto, M. Wouters, and V. Savona, Phys. Rev. B \textbf{77}, 125324 (2008).

\bibitem{Saito} H. Saito, T. Aioi, and T. Kadokura, Phys. Rev. Lett. \textbf{110}, 026401 (2013).

\bibitem{Grosso} G. Grosso, J. Graves, A. T. Hammack, A. A. High, L. V. Butov, M. Hanson, and A. C. Gossard, Nature Photon. \textbf{3}, 577 (2009).

\bibitem{oneModeNonl} I. Carusotto and C. Ciuti, Phys. Rev. Lett. \textbf{93}, 166401 (2004).

\bibitem{noLinCoupl} N. A. Gippius, I. A. Shelykh, D. D. Solnyshkov, S. S. Gavrilov, Y. G. Rubo, A. V. Kavokin, S. G. Tikhodeev, and G. Malpuech, Phys. Rev. Lett. \textbf{98}, 236401 (2007).

\bibitem{noCrossNonl} D. Sarchi, I. Carusotto, M. Wouters, and V. Savona, Phys. Rev. B \textbf{77}, 125324 (2008).

\bibitem{differentPump} S. S. Gavrilov, A. V. Sekretenko, S. I. Novikov, C. Schneider, S. H{\" o}fling, M. Kamp, A. Forchel, and V. D. Kulakovskii, Appl. Phys. Lett. \textbf{102}, 011104 (2013).

\bibitem{Tassone} F. Tassone and Y. Yamamoto, Phys. Rev. B \textbf{59}, 10830 (1999).

\bibitem{IXIXint} O. Kyriienko, E. B. Magnusson, and I. A. Shelykh, Phys. Rev. B \textbf{86}, 115324 (2012).

\bibitem{DXIXresonance} D. H. Auston, K. P. Cheung, J. A. Valdmanis, and D. A. Kleinman, Phys. Rev. Lett. \textbf{53}, 1555 (1984).

\bibitem{antenna} L. D. Landau and E. M. Lifshitz, \textit{The Classical Theory of Fields} (Butterworth-Heinemann, 1980).

\bibitem{Dicke} R. H. Dicke, Phys. Rev. \textbf{93}, 99 (1954).

\bibitem{superradianceNature} J. G. Bohnet, Z. Chen, J. M. Weiner, D. Meiser, M. J. Holland, and J. K. Thompson, Nature \textbf{484}, 7392 (2012).

\bibitem{de-Leon_2000} S. de-Leon and B. Laikhtman, Phys. Rev. B \textbf{61}, 2874 (2000).

\bibitem{de-Leon_2001} S. B. de-Leon and B. Laikhtman, Phys. Rev. B \textbf{63}, 125306 (2001).

\end{thebibliography}
\end{document}